\documentclass{elsart}
\usepackage{natbib}
\usepackage{graphicx}
\usepackage{amssymb}

\begin{document}

\begin{frontmatter}


\title{Quark clusters in quark stars and possible astrophysical implications}

\author{Weiwei Zhu, Renxin Xu}
\address{Astronomy Department, Peking University, Beijing, China 100871}

\begin{abstract}
\label{abs} A quark-cluster state, rather than the
color-super-conductivity state, may appear in matter with
low-temperature but high density, since the phase transition of
chiral symmetry broken and that of color-confinement could not
occur simultaneously. Such quark clusters might be stable in
strange quark matter. Quark stars would have temperatures to be
low enough to freeze by residual color interaction between the
clusters, and become then solid ones.
The charge numbers and the sizes of quark clusters, as well as the
residual interactions in-between are investigated. The solid state
properties of quark stars are constrained by observations.

\end{abstract}


\end{frontmatter}

\section{Introduction}
\label{intro}

Since E. Witten suggested that strange quark matter may be the
true ground state of color interaction system \citep{wit84}, it
has been of great interest to find such matter in reality.
There are hitherto many kinds of attempts, which can be classified
into two ways.
One is to detect directly strange quark matter on earth or in
space (e.g., the ground and the AMS-II observations of strangelets
in cosmic rays), the other is to study distant objects by means of
astrophysics (e.g., to identify quark stars with strangeness).
As for the nature of quark matter of strange stars, current
researchers are focusing on quark condensation in momentum space
(e.g., 2SC, CFL and LOFF states).
Recently it has been proposed alternatively that quarks in strange
quark matter with high density and low temperature may condense in
position space (i.e., forming quark clusters which are {\em not}
color-confined), and may be in a solid state when the temperature
is much low \citep{xu03,xu04}. This kind of solid quark stars
could be applied to explain well the discrepancy between the
observed glitches and free-precessions of pulsars
\citep{xu04,zhou04}.

In the following sections, based on the idea of quark clustering,
observational constrains on the nature of quark matter with quark
clusters are derived.

\section{Charge numbers and cluster sizes constrained by the featureless spectrum}
\label{elec}

The 500ks Chandra observation of RXJ1856.5-3754 \citep{fst1856}
shows a featureless Plank-like spectrum, which is not natural to
be explained in the atmosphere models of normal neutron star.
However, for the simplicity nature, the spectrum might be fitted
well in a solid quark star model \citep{zhang04}.
But if a cluster possesses more charges, it may behave like an
atom which should have complex energy levels and would thus show
corresponding atomic features in the spectrum.
Because the observational low-band of thermal X-ray spectra is
$\sim 0.1$ keV, this kind of ``atomic features'' can not be
observed if the charge number per cluster, $Z$, is not greater
than $\sim 2$.
In order to explain the observational fact of featureless spectra,
the number of quarks per cluster, $n$, can't be too large, because
$Z$ can be simply written as $n \cdot n_e/(3n_b)$.
In order to know the upper limit of $n$, we calculate electron
density, $n_e$, as well as baryon density $n_b$ near the surface
of the star.
As will be shown, these two parameters can be obtained by solving
the charge neutrality equation and state equation of zero pressure
\citep{alc86}, and are determined by three physical constant from
MIT bag model.

The charge neutrality equation is
\begin{equation}
\frac{2}{3}n_u - \frac{1}{3}n_d -\frac{1}{3}n_s - n_e=0,
\end{equation}
and the zero-temperature equation of state is
\begin{equation}
\frac{1}{3}(n_u+n_d+n_s)-\sum_i(\Omega_i+\mu_in_i)=B,
\end{equation}
where $i~(=u,~d,~s,~e)$ denotes up, down, and strange quarks, and
electron; $\Omega_i$, the thermal dynamic potentials; $\mu_i$, the
chemical potentials; and $n_i$, the number densities.
Here we take the formulae in the appendix of \citet{alc86} to
relate $\Omega_i$ and $\mu_i$,
\begin{equation}
\Omega_u=-\frac{\mu_u^4}{4\pi ^2}(1-\frac{2\alpha_s}{\pi}),
\end{equation}
\begin{equation}
\Omega_d=-\frac{\mu_d^4}{4\pi ^2}(1-\frac{2\alpha_s}{\pi}),
\end{equation}
\begin{equation}
\begin{array}{l}
\Omega_s=-\frac{1}{4\pi^2}(\mu_s(\mu_s^2-m_s^2)^{1/2}[\mu_s^2-\frac{5}{2}m_s^2]+\frac{3}{2}m_s^4\ln[\frac{\mu_s+(\mu_s^2-m_s^2)^{1/2}}{m_s}]\\
-\frac{2\alpha_s}{\pi}[3\{\mu-s(\mu_s^2-m_s^2)^{1/2}-m_s^2\ln[\frac{\mu_s+(\mu_s^2-m_s^2)^{1/2}}{\mu_s}]\}^2-2(\mu_s^2-m_s^2)^2\\
-m_s^4\ln^2\frac{m_s}{\mu_s}+6\ln\frac{\rho_R}{\mu_s}\{\mu_sm_s^2(\mu_s^2-m_s^2)^{1/2}-m_s^4\ln[\frac{\mu_s+(\mu_s+(\mu_s^2)-m_s^2)^{1/2}}{m_s}]\}]),
\end{array}
\end{equation}
\begin{equation}
\Omega_e=-\frac{\mu_e^4}{12\pi^2}.
\end{equation}
Beside eqs.(1)-(6), as a result of weak-interaction equilibriums
between quarks and electrons, the chemical potential $\mu$ of them
have the following relations,
\begin{equation}
\mu_d=\mu_s=\mu,
\end{equation}
\begin{equation}
\mu_u+\mu_e=\mu.
\end{equation}
We have also the relations between the number densities and the
thermodynamic potentials,
\begin{equation}
n_i=-\frac{\partial \Omega_i}{\partial\mu_i}.
\end{equation}

\begin{figure}[t]
  \centering
  \begin{minipage}[t]{.5\textwidth}
    \centering
    \includegraphics[width=7cm]{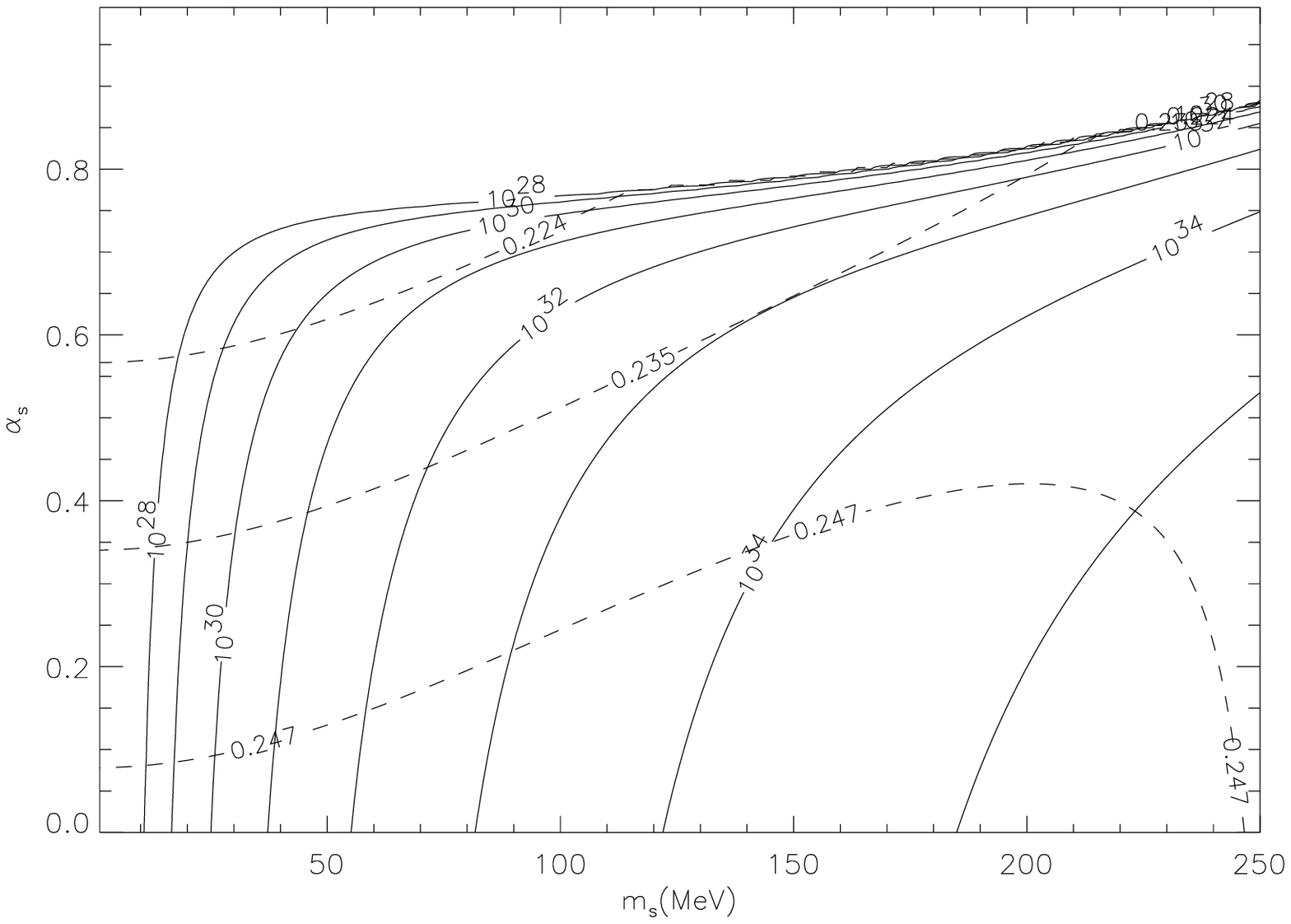}
  \end{minipage}%
  \begin{minipage}[t]{.5\textwidth}
    \centering
    \includegraphics[width=7cm]{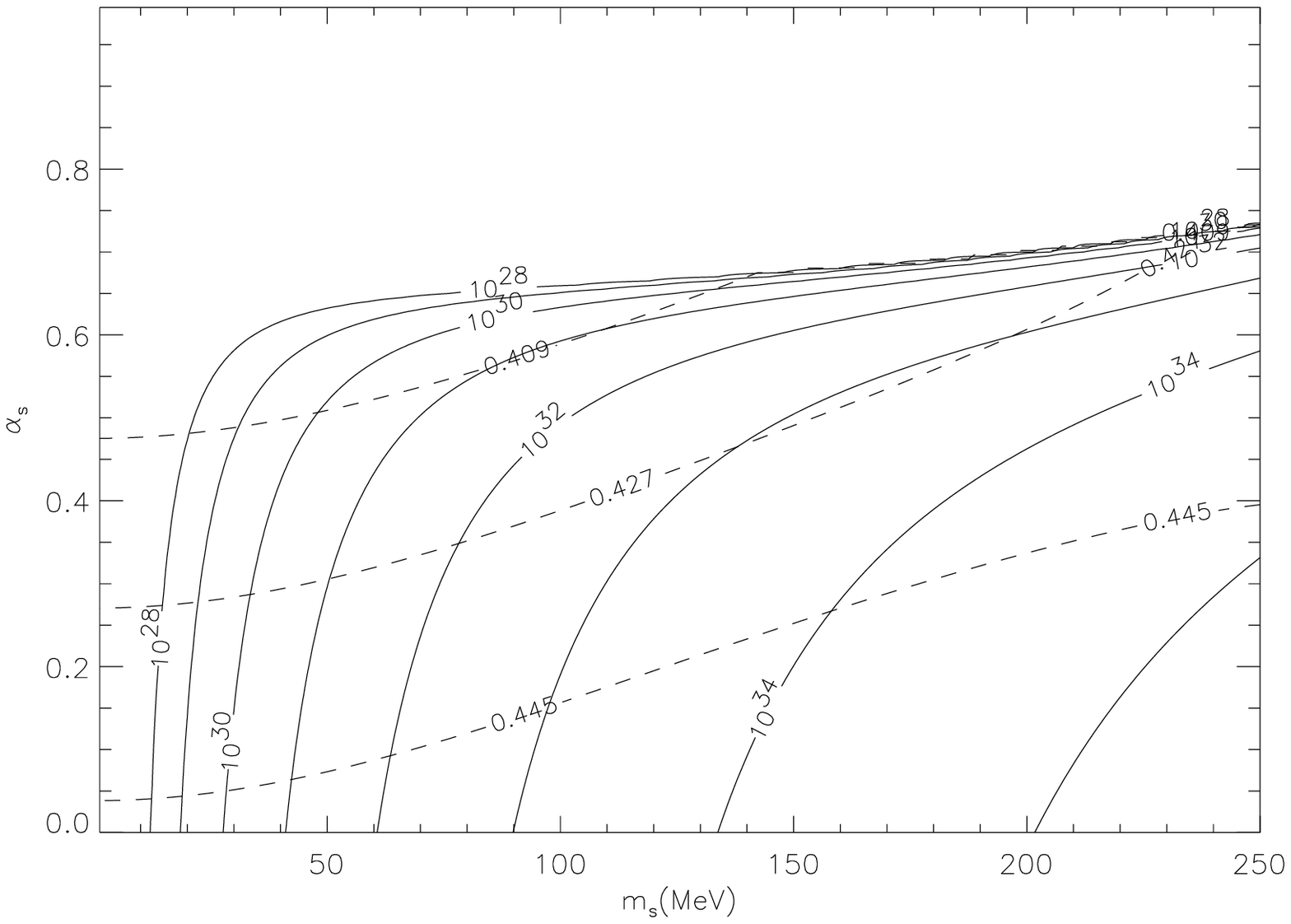}
  \end{minipage}%
\caption{Contour plots of electron density in unit of $\rm
cm^{-3}$ (solid lines) and of baryon number density in unit of
fm$^{-3}$ (dashed lines). The left panel is for $B^{1/4}=140\,\rm
MeV$, and the right for $B^{1/4}=170\,\rm MeV$.}
  \label{fig:ctb}
\end{figure}

Using above equations, the number densities of quarks and
electrons can be obtained, which are found to be dependent only on
three parameters of MIT bag model: the vacuum energy density $B$,
strange quark mass $m_s$, and strong interaction coupling constant
$\alpha_s$.
Our results show that for some value of these parameters the
electron density can drop to zero very fast. Beyond these
parameter values, no solution for $n_e$ exists any more.
Fig \ref{fig:ctb} shows the electron densities as parameters of
$m_s$ and $\alpha_s$ for different $B$. It is found that low $m_s$
and/or high $\alpha_s$ favor low $n_e$. This means that the
electric field occurred on the quark-star surface could be very
weak if $m_s$ is low and/or $\alpha_s$ is high.

\begin{figure}[t]
  \centering
  \begin{minipage}[t]{.5\textwidth}
    \centering
    \includegraphics[width=7cm]{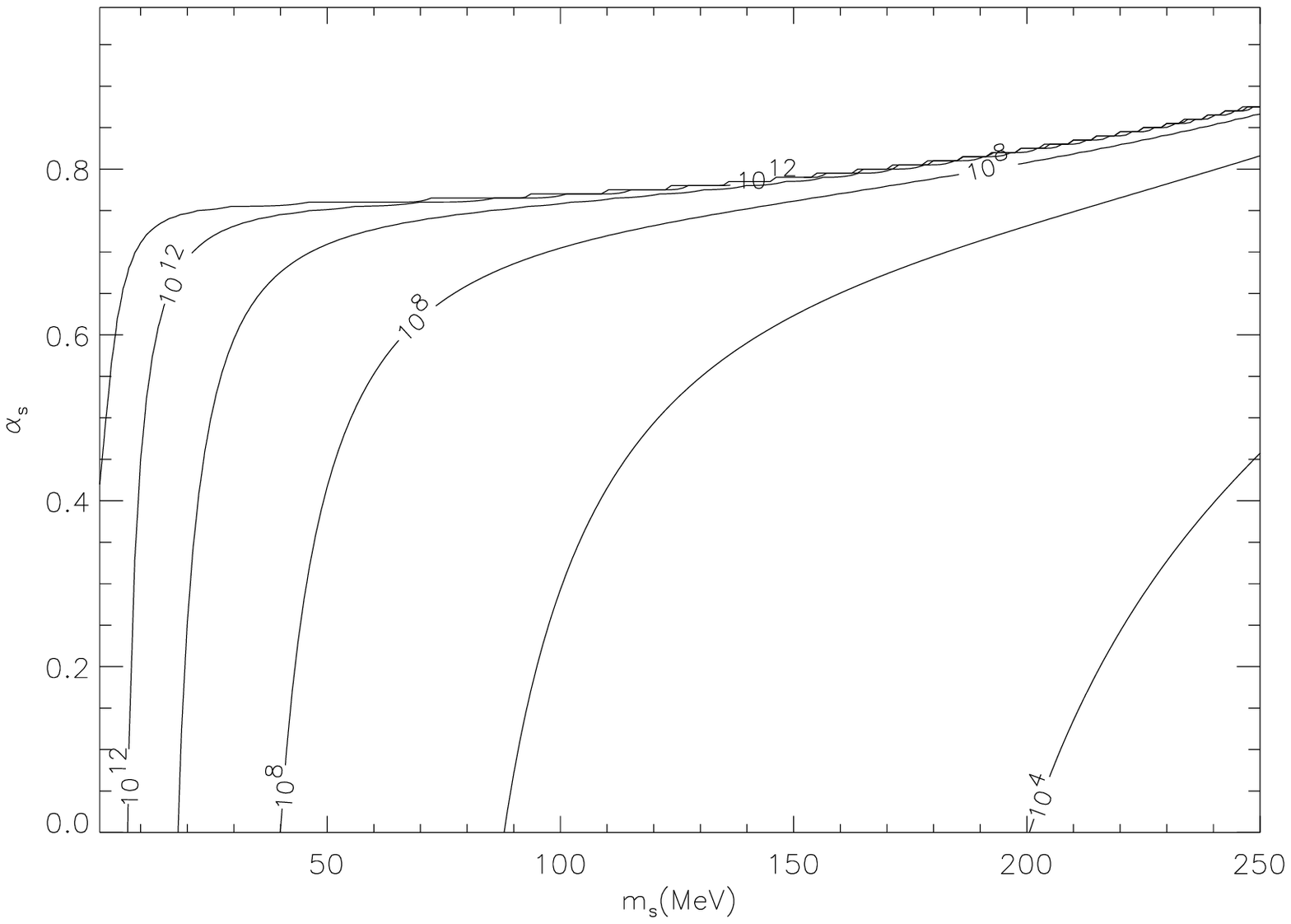}
  \end{minipage}%
  \begin{minipage}[t]{.5\textwidth}
    \centering
    \includegraphics[width=7cm]{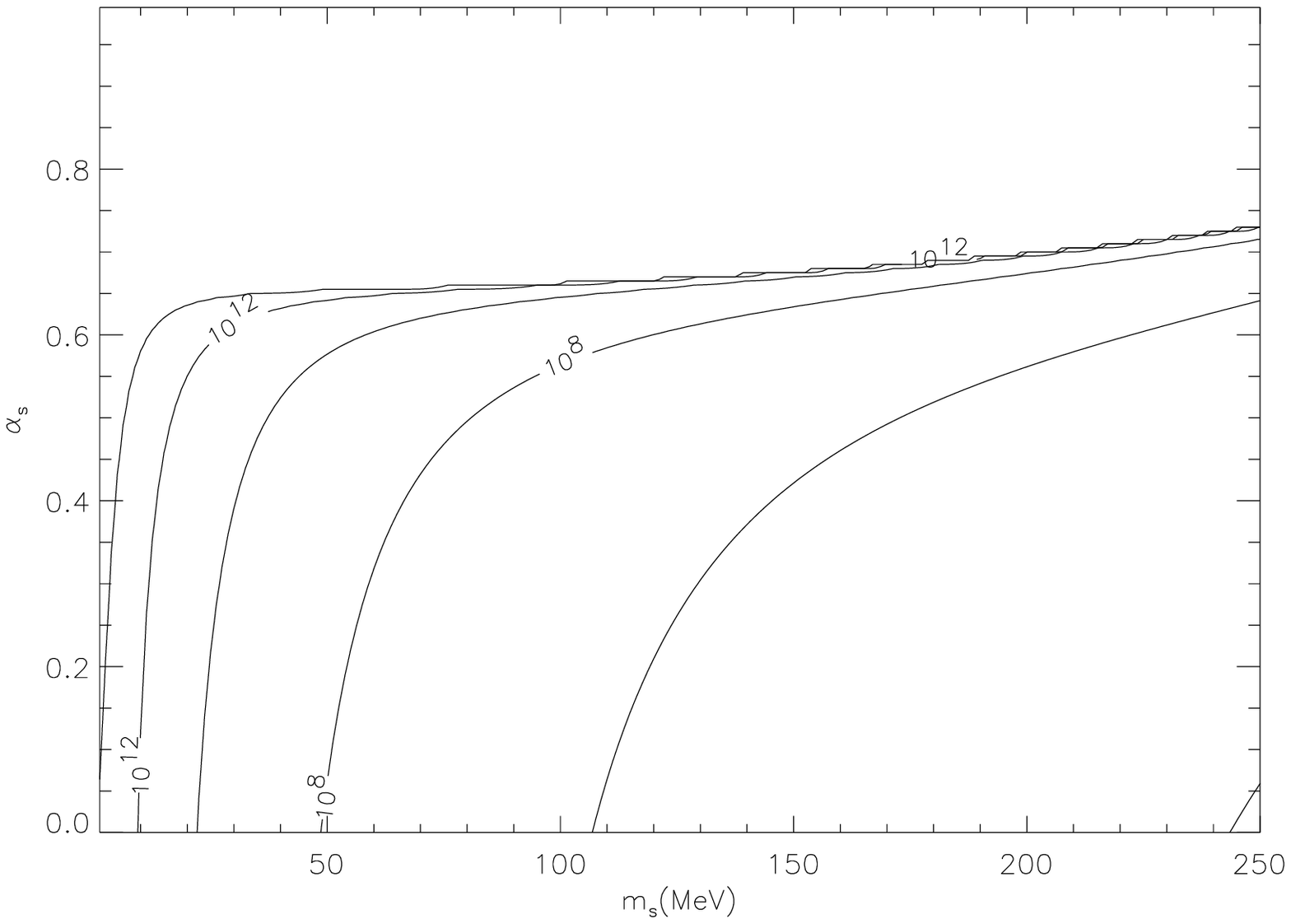}
  \end{minipage}%
\caption{The left panel shows the contour plot of $n_{\rm max}$
for $B^{1/4}=140\,\rm MeV$, and the right for $B^{1/4}=170\,\rm
MeV$.}
  \label{fig:ncl}
\end{figure}

With the charge density we now are able to constrain the number of
quarks per cluster $n$ from X-ray observation of RXJ1856.5-3754.
In order to reproduce the featureless thermal X-ray spectra, the
maximum value of quark number in a clusters, $n_{\rm max}$, is
calculated. Fig \ref{fig:ncl} shows the results of $n_{\rm max}$
for different parameters in the MIT bag model (we choose $Z=2$ in
the computations).
We find that the maximum quark number in clusters can vary in a
very large range of values.
For $50<m_s/{\rm MeV}<200$ and $0<\alpha_s<0.6$, we have
$10^4<n_{\rm max}<10^8$.

\section{Cluster size and interaction strength limited by glitches}
\label{glitch}

It was pointed out by \citet{zhou04} that if strange quark stars
are in a solid state, pulsar glitches could be understood well by
a sudden release of elastic energy accumulated when the shape of a
star changes as its spin slows down.
The elastic energy of a solid quark star is microscopically the
sum of bond-energies of all links between clusters, and is
determined by the strength of interactions and the sizes (i.e.,
$n$) of clusters.
Unfortunately up to now the strength of interaction can't be
calculated from first principles. We suppose thus its value to be
of the same order of the chemical potential of quarks; that is
within $\rm 1MeV-100MeV$.
%
%
By input different strengthes of interactions by hand, we can
obtain the magnitude of corresponding elastic energy.
%

A strange star should be in a solid state before a glitch occurs
due to increasing elastic energy.
There are generally two kinds of movement that are against the
solidification of strange stars; one is thermal motion, and the
other is the quantum uncertainty (i.e., the zero-point energy).
From the X-ray observation, we know that the thermal energy may be
at the magnitude of $0.1-1\rm KeV$. The zero-point energy can be
written as $\hbar^2/(2ml^2)$ if quark clusters are moving
non-relativistically, where $l\simeq [n/(3n_b)]^{1/3}$ is the
average distance between clusters.
From these expressions we can find that, for small clusters which
have less then thousands or hundreds of baryon numbers, the effect
of quantum uncertainty could be larger than that of thermal
motion.
This zero-point energy should be much smaller than the depth
($V_p$) of potential well between clusters,
$V_p\gg135n^{-5/3}n_b^{2/3}\,\rm MeV$, where $n_b$ is in unit of
$\rm fm^{-3}$ and $m\sim 300n$ MeV is suggested.
We have $V_p\gg 56 n^{-5/3}\, \rm MeV$ if $n_b=0.26 \rm fm^{-3}$
is chosen. This result means that, for $n\le 11$, the bond-energy
between clusters must be much higher than $1 \rm MeV$; otherwise
the matter should become a quantum fluid.
In an other word, if the interactions between clusters do exceed
much from $1 \rm MeV$, then it is natural for quark matter to be
solidified even when the clusters have only $\sim 10^2$ quarks.

To generate glitches, a solid quark star should have shear modulus
\citep{zhou04} $\mu\,\sim\,10^{30-34}\,\rm erg/cm^3$.
The shear modulus $\mu$ can be derived from the Young's modulus
$E$ via $\mu = E/[2(1+\nu)]$, where $\nu$ is called Poisson ratio,
$0<\nu<0.5$. We see $\mu$ and $E$ are of a same order of
magnitude.
For simplicity we can assume the potential of interaction is
Li\'{e}nard-Jones potential, $V(r)=-A/r^6+B/r^12$, and we have
then $l=(2B/A)^{1/6}$ and $V_p=-V(l)=A^2/(4B)$.
For finding a linear coefficient which cause the Hook's law, we
extract the Taylor expansion near the bottom of the potential
well,
\begin{equation}
V(r)=\frac{-A^2}{4\,B} + \frac{9}{2^{1/3}}
\frac{A^{7/3}}{B^{4/3}}[r-(\frac{2B}{A})^{1/6}]^2 +
  O[r-(\frac{2B}{A})^{1/6}]^3,
\end{equation}
from which we find that the second order term can be written as
$9/2^{1/3}\cdot A^{7/3}B^{-4/3}(r-l)^2 $.
Using $l$ and $V_p$ to eliminate $A$ and $B$, the second order
term becomes $36V_p(\Delta l/l)^2$.
At the same time, by considering a cylinder of length $L$ and
cross section $S$, the number of links between clusters at the
direction of length in this volume is $SL\cdot 3n_b/n$; thus the
value of elastic energy can be written as,
\begin{equation}
E_e\sim \frac{3n_b}{n}\,S\,L\,36V_p(\frac{\Delta
l}{l})^2=\frac{1}{2}Y(\frac{\Delta L}{L})^2.
\end{equation}
We can then easily get the Young's modulus $Y=72V_pn_b/n$.
The typical value of shear modulus is $ (1/3 \sim 1/2)Y$ in which
$n_b\sim0.26\,\rm fm^{-3}$, consequently,
\begin{equation}
\mu\sim 10^{34}\, n^{-1} (V_p/\rm MeV)~(\rm erg/cm^3).
\end{equation}
As mentioned before, from observations, we have shear modulus
$\mu\,\sim\,10^{30-34}$ erg/cm$^3$. The number of quarks per
cluster, $n$, has resultantly an upper limit, which is $10^4$ for
$V_p=1\rm MeV$ and $10^5$ for $V_p=10\rm MeV$.

\section{Conclusions and Discussions}

We have obtained the upper limits of the quark number per cluster,
$n$, from featureless spectra and pulsar glitches.
%
%
%
The maximum value of $n$, $n_{\rm max }$, constrained from
featureless spectra is $\sim 10^{4-8}$ for parameters of
$50<m_s/{\rm MeV}<200$, $0<\alpha_s<0.6$, and reasonable bag
constants $B$.
The upper limit of $n$ for generating pulsar glitches is $10^5$ if
bond-energy between clusters is 10MeV.
We conclude therefore that the upper limit of quark number per
cluster could be $\sim 10^{4-5}$ in order for bare strange stars
to reproduce both featureless X-ray spectra and pulsar glitches.

What have been done in this paper are based on the solid quark
star model.
This model is only a phenomenological model, into which it is very
necessary to research from first principles.
%
%
Nevertheless, the conclusion of our paper may provide helpful
information about the detail processes of quark clustering and
matter solidifying.

One problem in our paper is that we had used an equation of state
derived from the traditional concept of free-quark matter, not
clustered-quark matter.
This problem may affect the result significantly if clustering
changes the equation of state too much.
Although this issue needs to be treated seriously, the deviation
of equation of state may be slight, since color-confinement does
not occur yet in the quark-clustering matter and the property of
the vacuum could thus be not much different from that described by
\citet{alc86}.

\end{document}